\documentstyle[12pt,epsf]{article}
\begin{document}
\begin{center}
{\Large \bf Low-temperature expansion in link formulation. II}\\
\vspace*{1cm}
{\bf O.~Borisenko\footnote{email: oleg@ap3.bitp.kiev.ua}, \
V.~Kushnir\footnote{email: kushnir@ap3.bitp.kiev.ua}}\\

\vspace*{0.5cm}
{\large \it
N.N.Bogolyubov Institute for Theoretical Physics, National Academy
of Sciences of Ukraine, 252143 Kiev, Ukraine}\\
\vspace*{0.3cm}

\end{center}
\vspace{.5cm}

\begin{abstract}
We extend our previous analysis to arbitrary two dimensional  
$SU(N)$ principal chiral model in a link formulation. 
A general expression for the second order coefficient of fixed distance 
correlation function is given in terms of Green functions. 
This coefficient is calculated for distance 1 and is proven to be
path independent. We also study the weak coupling expansion of
the free energy of one dimensional $SU(N)$ model and explain
why it is non-uniform in the volume. Further,
we investigate the contribution of holonomy operators
to the low-temperature expansion in two dimensions.
All our results agree with the conventional expansion.
Nevertheless, we give some example which indicates
that the expansion in the link formulation can also
suffer from ambiguities previously found in the conventional 
perturbation theory.
\end{abstract}

\section{Introduction}

In our previous paper \cite{su2link} we proposed to use
an invariant link formulation to investigate some properties 
of two dimensional ($2D$) models in the low temperature limit. 
We have argued that this approach is more reliable for calculation
of asymptotic expansions of invariant functions in cases
when the Mermin-Wagner theorem forbids spontaneous symmetry
breaking in the thermodynamic limit. We have calculated 
the first order coefficient of the free energy of $SU(2)$
model which turned out to coincide with the conventional answer.
The starting point of calculations is the following partition function 
of $2D$ $SU(N)$ model 
\begin{equation}
Z = \int \prod_l dV_l
\exp \left[ \beta \sum_l {\mbox {Re Tr}} V_l + \ln J(V) \right] \ ,
\label{lPF}
\end{equation}
\noindent
where the Jacobian $J(V)$ is a product of $SU(N)$ delta-functions
taken over all plaquettes of $2D$ lattice
\begin{equation}
J(V) = \prod_p 
\left[ \sum_r d_r \chi_r \left( \prod_{l\in p}V_l \right) \right] \ .
\label{jacob}
\end{equation}
\noindent
The sum over $r$ is sum over all representations of $SU(N)$, 
$d_r=\chi_r(I)$ is the dimension of $r$-th representation. 
The $SU(N)$ character $\chi_r$ depends on a product of the link 
matrices $V_l$ along a closed path\footnote{We take plaquettes
for simplicity though one could choose any set of $L^2$
independent loops.}:
\begin{equation}
\prod_{l\in p}V_l = V_n(x)V_m(x+n)V_n^+(x+m)V_m^+(x) \ .
\label{prod}
\end{equation}
\noindent
One of the main technical observations made in \cite{su2link} is that
at large $\beta$ the $SU(N)$ delta-function can actually be replaced
by the Dirac delta-function so that partition function becomes
\begin{equation}
Z(\beta >> 1) = \int \prod_l dV_l
\exp \left[ \beta \sum_l {\mbox {Re Tr}} V_l \right] 
\prod_{p,k}\delta (\omega_p^k) \ ,
\label{lPFD}
\end{equation}
\noindent
where $\omega_p^k$ is a plaquette angle
\begin{equation}
\prod_{l\in p}V_l = \exp (i\lambda^k\omega_p^k) \ .
\label{plaqang}
\end{equation}
\noindent
Such a naive replacement is of course plagued by the problem of
$(N^2-1)$ zero modes for auxiliary fields therefore the delta-function 
in (\ref{lPFD}) should be correspondingly regularized. This can be done, for
instance, by introducing the heat-kernel into the expression for 
the partition function instead of the $SU(N)$ delta-function. 
This procedure is equivalent to introducing a mass term for
auxiliary fields. Direct calculations show however that zero modes
decouple from the large-$\beta$ expansion in the $SU(2)$ model \cite{su2link} 
(we could not generalize that consideration for arbitrary $SU(N)$, though).
In what follows we work with massless Green functions omitting zero modes
from all lattice sums similarly to $SU(2)$ case. All general expressions
remain valid if one works with mass regulator term as mentioned above.
Since all logarithmic divergences cancel we expect that the convergence
to the thermodynamic limit (TL) is uniform: in all cases the final result
can be expressed in terms of link functions $G_{ll^{\prime}}$,  
$D_l(x^{\prime})$ introduced in \cite{su2link} and standard $D$-functions.
For $SU(N)$ model the expansion itself is done precisely like in \cite{su2link}
for $SU(2)$. We obtain for the first order coefficient of the free energy
\begin{equation}
\frac{1}{2L^2}C^1_{f.e.} = \frac{(N^2-1)(N^2-2)}{64N} \ .
\label{C1fe}
\end{equation}
\noindent
This result agrees with the result of the conventional PT \cite{rossi}.

An essential motivation of our investigations is to study the problem
of the uniformity of the low-temperature expansion in $2D$ models,
the question addressed in \cite{superinst}. Before direct attack of this 
problem we have decided we need to make our approach more suitable
for practical computations. Also, it is necessary to establish some
general properties of the expansion, for example the path independence
of the correlation functions and the infrared finitness of the expansion.
We address these and some other problems in the present article.
We give a general expression for 
the second order coefficient of fixed distance 
correlation function in terms of Green functions and
calculate this coefficient for distance 1. 
Our final result agrees with Eq.(\ref{C1fe}).

Let us consider two-point correlation function 
\begin{equation}
\Gamma (x,y) = < \frac{1}{N}{\mbox {Re Tr}} \ U_xU_y^+ >=
< \frac{1}{N} {\mbox {Re Tr}} \prod_{l\in C_{xy}} W_l > \ ,
\label{corf}
\end{equation}
\noindent
where $W_l = V_l$ if along the path $C_{xy}$ the link $l$ goes in
the positive direction and $W_l = V_l^+$, otherwise.  
Obviously, it does not depend on a shape of the path $C_{xy}$
and any expansion in the link formulation must respect this property.
In large-$\beta$ expansion every coefficient at a given order
should be path independent. This requirement generates certain
relations on link Green functions thus allowing for a deformation
of any given path to any other one. Since this independence is guaranteed
only by the Jacobian $J(V)$, the check on such independence
would be highly desirable since it can show whether we treat
the Jacobian correctly and whether all contributions are taken 
into account. 

Further, we re-analyze the free energy of $1D$ models.
As is well known, large-$\beta$ expansion in $1D$ non-abelain models 
is non-uniform in the volume, therefore our expansion 
has to explain this feature.

As for now, all our results agree completely with the standard
approach to the low-temperature expansion in $2D$ models.
Nevertheless, we find that the expansion in the link
formulation also suffers from ambiguities previously found
in the standard expansion \cite{superinst}. Again, problem 
arises only for non-abelian models and only starting from
$1/\beta^2$. We mention this example in the Discussion.

This paper is organized as follows. In the next section we describe
some properties of link Green functions. 
In Section 3 we study the correlation function in $2D$ models.
We calculate the fixed distance correlation function in the $XY$ model.
The general expansion of the second order coefficient of
the correlation function in $SU(N)$ model is given in terms of Green functions.
In this paper we compute this coefficient for distance 1.
A proof of the path independence of the first and second order
coefficients of the correlation function is given in Section 4. 
Also, we prove here the infrared finitness
of the second order coefficient.
Section 5 is devoted to one dimensional models. 
We analyze the low-temperature expansion of the free energy 
and explain why this expansion is non-uniform in the volume. 
In Section 6 we study contribution of the holonomy operators
to the low-temperature expansion in $2D$ models and compare it with
one dimensional case. We summarize our results in Section 7.

\section{Link Green functions}

The main building blocks of the low-temperature expansion in the link
formulation are link Green functions $G_{ll^{\prime}}$ 
and $D_l(x^{\prime})$ introduced in \cite{su2link}. In this section
we describe some of their basic properties. 
The functions $G_{ll^{\prime}}$ and $D_l(x^{\prime})$
are defined as
\begin{equation}
G_{ll^{\prime}} = 2\delta_{l,l^{\prime}} - G_{x,x^{\prime}} -
G_{x+n,x^{\prime}+n^{\prime}} + G_{x,x^{\prime}+n^{\prime}} + G_{x+n,x^{\prime}} \ ,
\label{Gll1}
\end{equation}
\noindent
\begin{equation}
D_l(x^{\prime}) = G_{x,x^{\prime}} - G_{x+n,x^{\prime}} \ , 
\label{Dxl}
\end{equation}
\noindent
where link $l=(x,n)$ is defined by a point $x$ and a positive direction $n$.
$G_{x,x^{\prime}}$ is a ``standard'' Green function on the periodic lattice
\begin{equation}
G_{x,x^{\prime}} = \frac{1}{L^2} \sum_{k_n=0}^{L-1}
\frac{e^{\frac{2\pi i}{L}k_n(x-x^{\prime})_n}}
{f(k)} \ , \ k_n^2\ne 0 \ ,
\label{Gxx}
\end{equation}
\noindent
where we have denoted
\begin{equation}
f(k)=2-\sum_{n=1}^2\cos \frac{2\pi}{L}k_n \ .
\label{fk}
\end{equation}
Normalization is such that $G_{ll}=1$.
In the momentum space $G_{ll^{\prime}}$ reads
\begin{eqnarray}
G_{ll^{\prime}}=\frac{2\delta_{nn^{\prime}}-1}{L^2}
\sum_{k_n=0}^{L-1}\frac{e^{\frac{2\pi i}{L}k_n(x-x^{\prime})_n}}{f(k)}
F(n,n^{\prime}) + \frac{2\delta_{nn^{\prime}}}{L^2} \ , \nonumber  \\
F(n=n^{\prime})=2(1-\cos\frac{2\pi}{L}k_p) \ , n\ne p \ , \nonumber \\
F(n\ne n^{\prime})=(1-e^{\frac{2\pi i}{L}k_n})
(1-e^{-\frac{2\pi i}{L}k_{n^{\prime}}}) \ .
\label{Gllmom}
\end{eqnarray}
Using this representation it is easy to prove the following
``orthogonality'' relations for the link functions 
\begin{eqnarray}
\label{orthog1}
\sum_{b}G_{lb}G_{bl^{\prime}}=2G_{ll^{\prime}} \ ,   \\
\label{orthog2}
\sum_{b}D_b(x)G_{bl^{\prime}}=0 \ , \\
\label{orthog}
\sum_{b}D_b(x)D_b(x^{\prime})=2G_{x,x^{\prime}} \ ,
\end{eqnarray}
\noindent
where the sum over $b$ runs over all links of $2D$ lattice.
Let $C_{xy}$ be some path connecting points $x$ and $y$ and
let $C^d_{xy}$ be a path dual to the path $C_{xy}$, i.e consisting
of the dual links which are orthogonal to the original
links $l,l^{\prime}\in C_{xy}$. For simplicity let us consider
a path consisting of links which point only in a positive direction
(see Fig.1 for our notation of links on a dual lattice).
We then have
\begin{equation}
\sum_{l,l^{\prime}\in C^d_{xy}} G_{ll^{\prime}} = 2 D(x-y) \ ,
\label{GLXY1}
\end{equation}
\noindent
where 
\begin{equation}
D(x) = \frac{1}{L^2} \sum_{k_n=0}^{L-1}
\frac{1 - e^{\frac{2\pi i}{L}k_n x_n}}
{f(k)} \ , \ k_n^2\ne 0 .
\label{Dx}
\end{equation}
\noindent
\begin{figure}[t]
\centerline{\epsfxsize=4cm \epsfbox{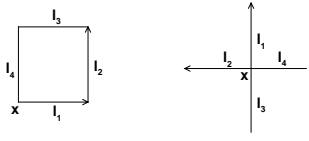}}
\hspace{0.5cm}
\caption{\label{dlink}Plaquette of original lattice and links of dual 
lattice. Link is determined by point $x$ and a positive
direction, e.g. $l_3=(x-n_1;n_1)$.}
\end{figure}
Let $\cal L$ be any closed path. Then
\begin{equation}
\sum_{l,l^{\prime}\in {\cal L}} \bar{G}_{ll^{\prime}} = 0 \ ,
\label{clpath}
\end{equation}
\noindent
where
$\bar{G}_{ll^{\prime}}=G_{ll^{\prime}}$ if both link $l$ and $l^{\prime}$
point in either positive or negative direction and
$\bar{G}_{ll^{\prime}}=-G_{ll^{\prime}}$ if one (and only one) of links 
points in negative direction. Let us demonstrate this property for
the simplest closed path, i.e. for a plaquette. 
From Fig.1 and definition (\ref{Gll1}) we have on 
the dual lattice\footnote{Numerical values for $G_{l_i,l_j}$ refer
to the thermodynamic limit.}
\begin{equation}
G_{l_1l_2}=G_{l_3l_4}=-G_{l_1l_4}=-G_{l_2l_3}=-1+\frac{2}{\pi} \ \ ,
G_{l_1l_3}=G_{l_2l_4}=-1+\frac{4}{\pi} \ .
\label{Gplaq}
\end{equation}
\noindent
We obtain from (\ref{clpath}) in case of plaquette
\begin{equation}
\sum_{l,l^{\prime}\in p}\bar{G}_{ll^{\prime}} = 
(4+8G_{l_1l_2}-4G_{l_1l_3}) = 0 \ .
\label{clplaq}
\end{equation}
\noindent
One sees that $G_{ll^{\prime}}$ satisfies the following equation
\begin{equation}
G_{l_1l^{\prime}}+G_{l_2l^{\prime}}-G_{l_3l^{\prime}}-
G_{l_4l^{\prime}} = 0
\label{Glleq}
\end{equation}
\noindent
for any link $l^{\prime}$. $D_l(x^{\prime})$ satisfies
the lattice Laplas equation
\begin{equation}
D_{l_1}(x^{\prime})+D_{l_2}(x^{\prime})-
D_{l_3}(x^{\prime})-D_{l_4}(x^{\prime})=2\delta_{x,x^{\prime}} \ .
\label{Dleq}
\end{equation}
\noindent
We shall also use the following equality 
\begin{equation}
\sum_x\left ( G_{l_1b_1}G_{l_2b_2}+G_{l_3b_1}G_{l_4b_2} - 
G_{l_1b_2}G_{l_2b_1}-G_{l_3b_2}G_{l_4b_1} \right ) = 0 \ ,
\label{Gllsum}
\end{equation}
\noindent
which is valid for any links $b_1,b_2$\footnote{On a finite lattice
equations (\ref{orthog1})-(\ref{orthog}) 
and (\ref{Glleq})-(\ref{Gllsum}) are valid
only up to $O(1/L^2)$ corrections.}.

\section{Correlation function in two-dimensional models}

Throughout this section we work on a dual lattice and follow 
notations of the previous section. Links $b_i$ denote links
belonging to the path $C^d_{xy}$.

\subsection{Correlation function in the $XY$ model}

In our previous paper we considered low-temperature expansion
of the free energy of the $XY$ model. Here we want to demonstrate
how conventional results for the fixed distance correlation function
can be recovered in the link formulation. We refer to \cite{su2link}
for expansion of the Gibbs measure and calculation of the generating
functional in the link formulation. 
For the fixed-distance correlation function we get
\begin{equation}
\Gamma_{XY} (x,y) = < \prod_{l\in C^d_{xy}} e^{i\phi_l} > =
1-\frac{1}{\beta}\Gamma^{(1)} (x,y)+\frac{1}{\beta^2}\Gamma^{(2)} (x,y)
+ O(\beta^{-3}) \ .
\label{XYcorf}
\end{equation}
\noindent
Coefficients $\Gamma^{(i)}$ are given by
\begin{equation}
\Gamma^{(1)} (x,y)=\frac{1}{4}\sum_{b_1b_2}G_{b_1b_2} \ ,
\label{XYC1}
\end{equation}
\noindent
\begin{equation}
\Gamma^{(2)} (x,y)=\frac{1}{32}
\left ( \sum_{b_1b_2}G_{b_1b_2} \right )^2 -
\ \frac{1}{32}\sum_{b_1b_2}\sum_lG_{b_1l}G_{lb_2} \ .
\label{XYC2}
\end{equation}
\noindent
Using orthogonality relation (\ref{orthog1}) and Eq.(\ref{GLXY1})
we obtain
\begin{equation}
\Gamma_{XY} (x,y) = 
1-\frac{1}{2\beta}D(x-y)+\frac{1}{8\beta^2}D(x-y)
\left [ D(x-y)-1 \right ] + O(\beta^{-3}) \ .
\label{XYcorres}
\end{equation}
\noindent
This result coincides with that quoted in the literature
for $O(2)$ model \cite{elitzur}.

\subsection{Correlation function in $SU(N)$ model: general expansion}

In this subsection we give general expression for the second-order
coefficient of the fixed-distance correlation function in terms
of Green functions. The generating functional is diagonal in group
indices and coincides with that given in \cite{su2link} for $SU(2)$.
Again, the expansion of the Gibbs measure and the Jacobian
is done precisely like in \cite{su2link}. Expanding (\ref{corf})
we write down
\begin{equation}
\Gamma_{SU(N)} (x,y) = 1-\frac{1}{\beta}\Gamma^{(1)} (x,y)+
\frac{1}{\beta^2}\Gamma^{(2)} (x,y) + ... \ .
\label{SUNcorf}
\end{equation}
\noindent
The first coefficient is given by
\begin{equation}
\Gamma^{(1)} (x,y)=\frac{N^2-1}{4N}\sum_{b_1b_2}G_{b_1b_2}=
\frac{N^2-1}{2N}D(x-y) 
\label{SUNC1}
\end{equation}
\noindent
and coincides with the conventional result.
We split the second coefficient into three pieces
\begin{equation}
\Gamma^{(2)} (x,y)=\frac{N^2-1}{16} \left ( Q_1+Q_2+Q_3 \right ) \ .
\label{SUNC2}
\end{equation}
\noindent
$Q_1$ describes contribution from the second order term of the correlation
function and zero order term of the Gibbs measure
\begin{equation}
Q_1=\frac{1}{N^2} \sum_{k=1}^4 Q_1^{(k)} \ ,
\label{Q1}
\end{equation}
\noindent
where
\begin{equation}
Q_1^{(1)}=\frac{2N^2-3}{6}\sum_b G^2_{bb} \ ,
\label{Q11}
\end{equation}
\noindent
\begin{equation}
Q_1^{(2)}=\sum_{b_1>b_2}\left [ (N^2-1)G_{b_1b_1} +
(N^2-2)G^2_{b_1b_2}+\frac{4}{3}(2N^2-3)G_{b_1b_2} \right ] \ ,
\label{Q12}
\end{equation}
\noindent
\begin{equation}
Q_1^{(3)}=\sum_{b_1\neq b_2\neq b_3}\left [ (N^2-1)G_{b_1b_2}+
(N^2-2)G_{b_1b_2}G_{b_2b_3} \right ] \ ,
\label{Q13}
\end{equation}
\noindent
\begin{equation}
Q_1^{(4)}=4\sum_{b_1>b_2>b_3>b_4}\left [ (N^2-1)
(G_{b_1b_2}G_{b_3b_4}+G_{b_1b_4}G_{b_2b_3})-
G_{b_1b_3}G_{b_2b_4} \right ] \ .
\label{Q14}
\end{equation}
\noindent
$Q_2$ describes contribution of $\beta^{-3/2}$ order from the
expansion of correlation function and of $\beta^{-1/2}$ order
from the expansion of the Jacobian. This ``self-connected''
piece is given by
\begin{eqnarray}
Q_2 = -\sum_{b_1>b_2>b_3}\sum_x\sum_{i<j}^4 
[ Q_x^{ij}(b_1,b_2,b_3)-Q_x^{ij}(b_1,b_3,b_2) + \nonumber  \\
Q_x^{ij}(b_2,b_3,b_1)-Q_x^{ij}(b_2,b_1,b_3) +
Q_x^{ij}(b_3,b_1,b_2)-Q_x^{ij}(b_3,b_2,b_1) ] \ ,
\label{Q2}
\end{eqnarray}
\noindent
where
\begin{equation}
Q_x^{ij}(b_1,b_2,b_3)= D_{b_1}(x)G_{b_2l_i}G_{b_3l_j} \ .
\label{Q22}
\end{equation}
\noindent
Finally, there are contributions of the first order terms from the
expansion of correlation function, the Gibbs measure and the Jacobian.
$Q_3$ describes the corresponding connected pieces
\begin{equation}
Q_3=\sum_{k=1}^3 Q_3^{(k)} \ ,
\label{Q3}
\end{equation}
\noindent
where
\begin{equation}
Q_3^{(1)}= \frac{1}{2N^2}\sum_{b_1b_2}\sum_lG_{b_1l}G_{lb_2} \ ,
\label{Q31}
\end{equation}
\noindent
\begin{eqnarray}
Q_3^{(2)}= - \frac{2}{3}\sum_{b_1b_2}\sum_x [ 3D_{b_2}(x)
(G_{b_1l_1}-G_{b_1l_4})(G_{l_1l_2}-G_{l_1l_3} ) + \nonumber \\ 
\frac{1}{2}\sum_{i=1}^4G_{b_1l_i}G_{b_2l_i}
(D_{l_1}(x) + D_{l_2}(x) - D_{l_3}(x) - D_{l_4}(x) ) + \nonumber  \\
(G_{b_1l_1}+G_{b_1l_2})(2D_{l_1}(x)G_{b_2l_2}-2D_{l_4}(x)G_{b_2l_3}-
D_{l_2}(x)G_{b_2l_1} + D_{l_3}(x)G_{b_2l_4}) + \nonumber  \\
(D_{l_1}(x)+D_{l_2}(x))G_{b_1l_3}G_{b_2l_4} - 
(D_{l_3}(x)+D_{l_4}(x))G_{b_1l_1}G_{b_2l_2} ] \ ,
\label{Q32}
\end{eqnarray}
\noindent
\begin{equation}
Q_3^{(3)}= \frac{1}{4}\sum_{b_1b_2}\sum_{x,x^{\prime}}
\sum_{i<j}^4\sum_{i^{\prime}<j^{\prime}}^4
\left (G_{x,x^{\prime}}I_1 + I_2 \right ) \ .
\label{Q33}
\end{equation}
\noindent
We have denoted
\begin{eqnarray}
I_1=G_{b_1,l_i}G_{b_2,l^{\prime}_{i^{\prime}}}
G_{l_j,l^{\prime}_{j^{\prime}}} +
G_{b_1,l_j}G_{b_2,l^{\prime}_{j^{\prime}}}
G_{l_i,l^{\prime}_{i^{\prime}}} - \nonumber  \\
G_{b_1,l_i}G_{b_2,l^{\prime}_{j^{\prime}}}
G_{l_j,l^{\prime}_{i^{\prime}}} -
G_{b_1,l_j}G_{b_2,l^{\prime}_{i^{\prime}}}
G_{l_i,l^{\prime}_{j^{\prime}}} \ ,
\label{I1}
\end{eqnarray}
\noindent
\begin{eqnarray}
I_2=D_{b_1}(x)D_{b_2}(x^{\prime})(G_{l_i,l^{\prime}_{j^{\prime}}}
G_{l_j,l^{\prime}_{i^{\prime}}}-G_{l_i,l^{\prime}_{i^{\prime}}}
G_{l_j,l^{\prime}_{j^{\prime}}}) + \\
2 D_{b_1}(x)D_{l_i}(x^{\prime})(G_{b_2,l^{\prime}_{i^{\prime}}}
G_{l_j,l^{\prime}_{j^{\prime}}}-G_{b_2,l^{\prime}_{j^{\prime}}}
G_{l_j,l^{\prime}_{i^{\prime}}}) +\nonumber  \\
2 D_{b_1}(x)D_{l_j}(x^{\prime})(G_{b_2,l^{\prime}_{j^{\prime}}}
G_{l_i,l^{\prime}_{i^{\prime}}}-G_{b_2,l^{\prime}_{i^{\prime}}}
G_{l_i,l^{\prime}_{j^{\prime}}}) + \nonumber  \\
D_{l^{\prime}_{i^{\prime}}}(x)G_{b_1,l^{\prime}_{j^{\prime}}}
(D_{l_i}(x^{\prime})G_{b_2,l_j}-
D_{l_j}(x^{\prime})G_{b_2,l_i}) + \nonumber  \\
D_{l^{\prime}_{j^{\prime}}}(x)G_{b_1,l^{\prime}_{i^{\prime}}}
(D_{l_j}(x^{\prime})G_{b_2,l_i}-
D_{l_i}(x^{\prime})G_{b_2,l_j}) \ . \nonumber
\label{I2}
\end{eqnarray}
\noindent
In all formulae link $l_i$ ($l^{\prime}_{j^{\prime}}$) refers to one 
of four links attached to a given site $x$ ($x^{\prime}$) (see Fig.1).
As it stays, this expression for the second order coefficient
is valid for any path if all links $b_i\in C^d_{xy}$ 
point in positive directions. If one considers a path where 
some links point in a negative direction, one has to change
a sign of the corresponding Green functions.

\subsection{$\Gamma_{SU(N)} (1,0)$}

Here we compute correlation function for distance 1. 
From (\ref{SUNcorf})-(\ref{SUNC2}) we have
\begin{equation}
\Gamma_{SU(N)} (1,0) = 1-\frac{1}{\beta}\frac{N^2-1}{4N}+
\frac{1}{\beta^2}\frac{N^2-1}{16} \left ( Q_1+Q_3 \right ) \ ,
\label{SUNcor1}
\end{equation}
\noindent
where we took into account that $D(1,0)=1/2$ and $Q_2=0$ in this case.
Since only $Q_1^{(1)}$ contributes to $Q_1$ we easily find
\begin{equation}
Q_1=\frac{2N^2-3}{6N^2} \ .
\label{Q1res}
\end{equation}
\noindent
To calculate $Q_3$ we first note that one has to put $b_1=b_2=b$,
where $b$ is any fixed link. 
From (\ref{Q31}) we obtain using (\ref{orthog1})
\begin{equation}
Q_3^{(1)}= \frac{1}{N^2} \ .
\label{Q31res}
\end{equation}
\noindent
To calculate $Q_3^{(2)}$ we make use the translation invariance
to write down
\begin{equation}
Q_3^{(2)}= \frac{1}{2L^2}\sum_b Q_3^{(2)} \ . 
\label{Q32tr}
\end{equation}
\noindent
Using again (\ref{orthog1})-(\ref{orthog2}) we find 
\begin{equation}
Q_3^{(2)}= - \frac{2}{3}\left (3+2G_{l_1l_3}-G_{l_1l_2} \right )=
- \frac{4}{3}\left (1+\frac{3}{\pi} \right ) \ .
\label{Q32res}
\end{equation}
\noindent
The same tricks applied to $Q_3^{(3)}$ give
\begin{equation}
Q_3^{(3)}=\frac{1}{2}\left (B_1+B_2 \right ) \ , 
\label{Q33res}
\end{equation}
\noindent
where
\begin{equation}
B_1 = \frac{1}{2L^2}\sum_{x,x^{\prime}}
\sum_{i<j}^4\sum_{i^{\prime}<j^{\prime}}^4
G_{x,x^{\prime}} 
(G_{l_i,l^{\prime}_{i^{\prime}}}G_{l_j,l^{\prime}_{j^{\prime}}} -
G_{l_i,l^{\prime}_{j^{\prime}}}G_{l_j,l^{\prime}_{i^{\prime}}}) \ ,
\label{B1}
\end{equation}
\noindent
\begin{eqnarray}
B_2=\frac{1}{2L^2}\sum_{x,x^{\prime}}
\sum_{i<j}^4\sum_{i^{\prime}<j^{\prime}}^4
(G_{l_i,l^{\prime}_{i^{\prime}}}D_{l^{\prime}_{j^{\prime}}}(x) 
D_{l_{j}}(x^{\prime}) + 
G_{l_j,l^{\prime}_{j^{\prime}}}D_{l^{\prime}_{i^{\prime}}}(x) 
D_{l_{i}}(x^{\prime}) -  \nonumber  \\
G_{l_i,l^{\prime}_{j^{\prime}}}D_{l^{\prime}_{i^{\prime}}}(x) 
D_{l_{j}}(x^{\prime}) - 
G_{l_j,l^{\prime}_{i^{\prime}}}D_{l^{\prime}_{j^{\prime}}}(x) 
D_{l_{i}}(x^{\prime})) \ . 
\label{B2}
\end{eqnarray}
\noindent
$B_1$ and $B_2$ have been calculated in \cite{su2link} and equal
\begin{equation}
B_1 = \frac{1}{2} \ \ , \ \ B_2 = \frac{8}{\pi}+1 \ .
\label{B1B2res}
\end{equation}
\noindent
We thus find
\begin{equation}
Q_1+Q_3 = - \frac{N^2-2}{4N^2} \ .
\label{Q1Q3}
\end{equation}
\noindent
It leads to a final result
\begin{equation}
\Gamma_{SU(N)} (1,0) = 1-\frac{1}{\beta}\frac{N^2-1}{4N}-
\frac{1}{\beta^2}\frac{(N^2-1)(N^2-2)}{64N^2}  \ ,
\label{SUNcorres}
\end{equation}
\noindent
which agrees with result for the free energy, Eq.(\ref{C1fe}).

\section{Path independence of $\Gamma_{SU(N)} (x,y)$}

We want to prove now that our representation for the correlation function 
is path independent. In the strong coupling region
such an independence is a simple consequence of the compact
invariant integration which is lost in the large-$\beta$ expansion.   
Since, however any coefficient at a given order of the large-$\beta$ 
expansion has to be path independent the direct proof of this property
serves as an additional check of the correctness of the expansion.
The path independence is guaranteed by the Jacobian which introduces
the constraint on the plaquette matrix $V_p=1$. Just this constraint
imposes independence of the correlation function of the path
and this constraint must be checked at each order of the large-$\beta$ 
expansion. To show that this constraint is fulfilled we write down
\begin{equation}
\Gamma_p = < \frac{1}{N} {\mbox {Re Tr}} V_p > =
1-\frac{1}{\beta}\Gamma_p^{(1)}+\frac{1}{\beta^2}\Gamma_p^{(2)} \ .
\label{Gpl}
\end{equation}
\noindent
The path independence requires
$$
\Gamma_p^{(1)}=\Gamma_p^{(2)}=0 \ .
$$
First, we note that the expansion for the correlation function
given in section 3.2 remains valid also for $\Gamma_p$ 
where one has to identify links $b_i$ with links belonging to a
given plaquette as shown in Fig.1. Besides, one has to change
a sign in some Green functions. If we write
$$
V_p=V_{l_1}V_{l_2}V^+_{l_3}V^+_{l_4}
$$ 
the signs should be changed in $G_{ll_3}$, $G_{ll_4}$,
$D_{l_3}(x)$ and $D_{l_4}(x)$.

The equality $\Gamma_p^{(1)}=0$ trivially follows from
(\ref{clpath}). We parametrize $\Gamma_p^{(2)}$ as
\begin{equation}
\Gamma_p^{(2)} = \frac{N^2-1}{16} 
\left ( \frac{1}{N^2}C_1+C_2 \right ) \ .
\label{G2par}
\end{equation}
\noindent
Since $N$ is arbitrary, the coefficients $C_1$ and $C_2$ should
be equal to zero independently. 
There are two contributions to $C_1$ coming from $Q_1$ and
from $Q_3^{(1)}$. We find via direct computation
\begin{eqnarray}
C_1 = -\frac{1}{2}\sum_b G^2_{bb}-\sum_{b_1>b_2}
\left [ 1 + 2G^2_{b_1b_2}+ 4G_{b_1b_2} \right ] - \nonumber \\
2\sum_{b_1>b_2>b_3}\left [ G_{b_1b_2}+G_{b_2b_3}+G_{b_1b_3} +
2(G_{b_1b_2}G_{b_2b_3}+G_{b_1b_3}G_{b_3b_2}+
G_{b_1b_2}G_{b_1b_3}) \right ] - \nonumber  \\
4\sum_{b_1>b_2>b_3>b_4}\left [ G_{b_1b_2}G_{b_3b_4}+G_{b_1b_4}G_{b_2b_3}+
G_{b_1b_3}G_{b_2b_4} \right ] + \nonumber \\
\frac{1}{2}\sum_{b_1b_2}\sum_lG_{b_1l}G_{lb_2} = 0 .
\label{C1Gp}
\end{eqnarray}
\noindent
$C_2$ consists of the following terms 
\begin{equation}
C_2=Q_1(p)+Q_2(p)+Q_3^{(2)}(p)+Q_3^{(3)}(p) \ ,
\label{C2def}
\end{equation}
\noindent
where we have introduced obvious definitions.
For $Q_1(p)$ (terms at $N^2$ in (\ref{Q11})-(\ref{Q14})) we get
\begin{eqnarray}
Q_1(p)= \frac{1}{3}\sum_b G^2_{bb}+\sum_{b_1>b_2}
\left [ 1 + G^2_{b_1b_2}+\frac{8}{3} G_{b_1b_2} \right ] + \nonumber \\
2\sum_{b_1>b_2>b_3}\left [ G_{b_1b_2}+G_{b_2b_3}+G_{b_1b_3} +
G_{b_1b_2}G_{b_2b_3}+G_{b_1b_3}G_{b_3b_2}+
G_{b_1b_2}G_{b_1b_3} \right ] + \nonumber  \\
4\sum_{b_1>b_2>b_3>b_4}
\left [ G_{b_1b_2}G_{b_3b_4}+G_{b_1b_4}G_{b_2b_3} \right ] = 
-4 \left (1+G_{l_1l_2} \right )^2 = -\frac{16}{\pi^2} \ .
\label{Q1p}
\end{eqnarray}
\noindent
$Q_2(p)$ is given in (\ref{Q2}) and equals zero for any closed
path. This can be most easily seen from the cyclic permutations of links
under the trace in (\ref{Gpl}). From (\ref{Glleq}) follows that
$$
Q_3^{(2)}(p)=0
$$
and 
\begin{equation}
Q_3^{(3)}(p) = \frac{1}{4}\sum_{b_1b_2}\sum_{x,x^{\prime}}
\sum_{i<j}^4\sum_{i^{\prime}<j^{\prime}}^4 I_2(p) \ .
\label{Q33pl}
\end{equation}
\noindent
Only one term survives the summation over $b_i$. Eq.(\ref{Dleq})
gives
\begin{equation}
Q_3^{(3)}(p) = \sum_{x,x^{\prime}}\delta_{x,0}\delta_{x^{\prime},0}
\sum_{i<j}^4\sum_{i^{\prime}<j^{\prime}}^4 
(G_{l_i,l^{\prime}_{j^{\prime}}}
G_{l_j,l^{\prime}_{i^{\prime}}}-G_{l_i,l^{\prime}_{i^{\prime}}}
G_{l_j,l^{\prime}_{j^{\prime}}})  \ .
\label{Q33plres}
\end{equation}
\noindent
Performing all summations we finally get
\begin{equation}
Q_3^{(3)}(p) = 4 \left (1+G_{l_1l_2} \right )^2 = 
\frac{16}{\pi^2} \ .
\label{Q33plcontr}
\end{equation}
\noindent
One sees that this connected piece cancels exactly contribution
of the second order term from the expansion of the correlation function.
We get convinced that
$$
\Gamma_p^{(2)}=0 \ .
$$
As a matter of fact, Eqs.(\ref{clpath}), (\ref{Glleq}) and (\ref{Dleq}) 
are short-hand statements of the independence of the correlation function
of the path. Using these equations one deforms any path expressing
some link from a given plaquette through remaining ones.
The generalization of this proof to the product of link matrices
along an arbitrary closed path is straightforward.

\subsection{Infrared finitness of $\Gamma^{(2)} (x,y)$}

In this subsection we prove the infrared finitness of 
$\Gamma^{(2)} (x,y)$. The only term which includes 
$G_{x,x^{\prime}}$ and hence could logarithmically diverge is 
\begin{equation}
Q_3^{(div)} = \sum_{b_1b_2}\sum_{x,x^{\prime}}
\sum_{i<j}^4\sum_{i^{\prime}<j^{\prime}}^4 G_{x,x^{\prime}}I_1 \ 
\label{Q33div}
\end{equation}
\noindent
and $I_1$ is given in Eq.(\ref{I1}).
We want to show that $G_{x,x^{\prime}}$ in the last expression can actually
be replaced by $G_{x,x^{\prime}}-G_0$ because all additive constants like
logarithmic term in $G_{x,x^{\prime}}$ do not contribute to the sums
in Eq.(\ref{Q33div}).
Thus, one has to prove that
\begin{equation}
S=\sum_{b_1b_2}\sum_{x,x^{\prime}}
\sum_{i<j}^4\sum_{i^{\prime}<j^{\prime}}^4 I_1 = 0 \ . 
\label{canc}
\end{equation}
\noindent
Indeed, using (\ref{orthog1}) $S$ can be transformed to
\begin{equation}
S=\frac{1}{2}\sum_{b_1b_2}\sum_{b}
\left [ A(b_1b)-A(bb_1) \right ]
\left [ A(b_2b)-A(bb_2) \right ] \ ,
\label{Str}
\end{equation}
\noindent
where sum over $b$ runs over all links of the lattice and
\begin{equation}
A(b_1b)=\sum_x \sum_{i<j}^4G_{l_ib_1}G_{l_jb} \ .
\label{Atr}
\end{equation}
\noindent
The equality
\begin{equation}
A(b_1b)-A(bb_1) = 0 
\label{Atreq0}
\end{equation}
\noindent
follows from (\ref{Glleq}) and (\ref{Gllsum}). Therefore, $S=0$
and Eq.(\ref{Q33div}) is infrared finite.

\section{One-dimensional models in the link formulation}

To analyze $1D$ $SU(N)$ model we note that the formula for 
$\Gamma^{(2)} (x,y)$ given in section 3.2 remains valid
if we take for the link Green function the following expression
\begin{equation}
G_{ll^{\prime}} = 2\delta_{l,l^{\prime}} \ .
\label{Gll1D}
\end{equation}
\noindent
This equation is a trivial consequence of the fact
that in the link formulation the $1D$ model reduces to one link integrals.
Then, it is straightforward to calculate, for example, the first order
coefficient of the free energy. We find
\begin{equation}
\frac{1}{L}C^1_{f.e.} = - \frac{N^2-1}{8N} \ ,
\label{C1fe1D}
\end{equation}
\noindent
which agrees with the expansion of the exact result in the TL.
On the other hand, it is well known that the low-temperature expansion
in $1D$ non-abelian models is non-uniform in the volume, in particular
the conventional PT produces result different from the Eq.(\ref{C1fe1D}).
To explain this non-uniformity we remind that on the periodic lattice
one has to constraint a holonomy operator if one works in the link formulation.
Working with $2D$ models we have neglected this additional constraint since
it seems to us rather unlikely that such global constraint may influence
the TL in $2D$ (see next section for details). 
This happens, however in $1D$ model as we are going to show below.

The partition function is given by
\begin{equation}
Z = \int \prod_l dV_l
\exp \left[ \beta \sum_l {\mbox {Re Tr}} V_l + \ln J(V) \right] \ ,
\label{lPF1D}
\end{equation}
\noindent
where $J(V)$ introduces a global constraint on link matrices
\begin{equation}
J(V) = \sum_r d_r \chi_r \left( \prod_{l=1}^LV_l \right) \ .
\label{jacob1D}
\end{equation}
\noindent
Again, at large $\beta$ we replace the $SU(N)$ delta-function 
by the Dirac delta-function, i.e.
\begin{equation}
J(V) = \int\prod_{k=1}^{N^2-1}d\phi_k\exp [-i\phi_k\omega^k(C)] \ ,
\label{j1DD}
\end{equation}
\noindent
where $\omega^k(C)$ is defined as
\begin{equation}
\prod_{l=1}^LV_l = \exp [i\lambda^k\omega^k(C)] \ .
\label{Cang}
\end{equation}
\noindent
We omit all technical details which are exactly the same as in $2D$.
For the first order coefficient of the free energy we find in 
the large volume limit
\begin{equation}
\frac{1}{L}C^1_{SU(N)} = - \frac{N^2-1}{8N}+\frac{N(N^2-1)}{24} \ .
\label{C11D}
\end{equation}
\noindent
The second term on the right-hand side of the last formula comes from
the expansion of $J(V)$ and modifies the correct expression
(\ref{C1fe1D}).
Our result (\ref{C11D}) disagrees with the one given in \cite{rossi}.
We think it is because the result of \cite{rossi} was obtained using
the mass regulator term, i.e. the procedure which is known to give wrong 
answer even in a finite volume \cite{ug}. To check the correctness
of (\ref{C11D}) we have compared it for $N=2$ with $O(n=4)$ model 
\begin{equation}
\frac{1}{L}C^1_{O(n)} = \frac{n-1}{8}-\frac{(n-1)(n-2)}{24} \ ,
\label{C11DOn}
\end{equation}
\noindent
where the second term comes from the Hasenfratz term which survives 
the TL in $1D$. One sees that results indeed coincide\footnote{One needs
also to replace $\beta\to 2\beta$ in $O(4)$.}.

\section{Holonomy operators in $2D$}

On $2D$ lattice one should restrict two holonomy operators.
In our previous analysis we have neglected this restriction.
Since, however this global constraint influences the TL in $1D$
if the low-temperature expansion is done in a finite volume
we think it is instructive to see what happens with holonomies
in two-dimensional models. 

Let $H_i (i=1,2)$ be any given path winding around the whole lattice.
$H_1$ and $H_2$ are orthogonal to each other. One has to introduce
two global constraints into the partition function (\ref{lPF})
\begin{equation}
J(H) = \int\prod_{k=1}^{N^2-1}\prod_{i=1}^2
d\phi_k(i)\exp [-i\phi_k(i)\omega^k(H_i)] \ ,
\label{j2DD}
\end{equation}
\noindent
where $\omega^k(H_i)$ is defined as
\begin{equation}
\prod_{l\in H_i}V_l = \exp [i\lambda^k\omega^k(H_i)] \ .
\label{Cang2D}
\end{equation}
\noindent
There are two types of contributions from $J(H)$. The first one
comes from the expansion of $J(H)$ itself. It is too cumbersome
to be given here in full. This contribution, however can be 
expressed only through link Green functions and is proportional
to the linear size of the lattice. This is a reason why it
survives the TL in $1D$. Correspondingly, in $2D$ it vanishes
like $O(1/L)$. The second type is related to the modification
of the generating functional. Namely, one should make the
following replacement of the Green functions 
\begin{equation}
G_{ll^{\prime}} \to \bar{G}_{ll^{\prime}} = G_{ll^{\prime}} - 
\frac{1}{2}\sum_i\sum_{(bb^{\prime})\in H_i}G_{lb}
G_{l^{\prime}b^{\prime}} \ ,
\label{GllH}
\end{equation}
\noindent
\begin{equation}
D_l(x) \to \bar{D}_l(x) = D_l(x) - 
\frac{1}{2}\sum_i\sum_{(bb^{\prime})\in H_i}D_b(x)
G_{lb^{\prime}} \ ,
\label{DlH}
\end{equation}
\noindent
\begin{equation}
G_{x,x^{\prime}} \to \bar{G}_{x,x^{\prime}} = G_{x,x^{\prime}} + 
\frac{1}{2}\sum_i\sum_{(bb^{\prime})\in H_i}D_b(x)
D_{b^{\prime}}(x^{\prime}) \ .
\label{GxxH}
\end{equation}
\noindent
In particular, the corresponding replacements should be made
in formulae (\ref{SUNcorf})-(\ref{I2}).
Let us take for simplicity such paths $H_i$ which consist
of links pointing only in one direction. The coordinates of the 
corresponding links on the dual lattice are
$$
b=(x_1,0;n_2) \ x_1\in [0,L-1] \ , \ b\in H_1
$$
and
$$
b=(0,x_2;n_1) \ x_2\in [0,L-1] \ , \ b\in H_2 \ .
$$
It is easy to find
\begin{equation}
\sum_{b\in H_1}G_{lb}=0 \ , \ l=(x,n_1) \ ,
\label{sumn1GllH}
\end{equation}
\noindent
\begin{equation}
\sum_{b\in H_1}G_{lb}=\frac{2}{L} \ , \ l=(x,n_2) \ .
\label{sumn2GllH}
\end{equation}
\noindent
In general we get for $l=(x,n)$, $l^{\prime}=(x^{\prime},n^{\prime})$
\begin{equation}
\frac{1}{2}\sum_i\sum_{(bb^{\prime})\in H_i}G_{lb}
G_{l^{\prime}b^{\prime}} = \frac{2}{L^2}\delta_{nn^{\prime}} \ .
\label{GllHf}
\end{equation}
\noindent
On the other hand, one finds in the TL
\begin{eqnarray}
\sum_{b\in H_1}D_b(x)=2\delta_{x_2,0}-1 \ , \nonumber  \\
\sum_{b\in H_2}D_b(x)=2\delta_{x_1,0}-1 \ ,
\label{sumDlH}
\end{eqnarray}
\noindent
which leads to
\begin{equation}
\frac{1}{2}\sum_i\sum_{(bb^{\prime})\in H_i}D_b(x)
D_{b^{\prime}}(x^{\prime})=\frac{1}{2}\sum_{n=1}^2
(2\delta_{x_n,0}-1)(2\delta_{x^{\prime}_n,0}-1) \ .
\label{sumDDH}
\end{equation}
\noindent
We thus have for new functions
\begin{equation}
\bar{G}_{ll^{\prime}} = G_{ll^{\prime}} - O(1/L^2) \ ,
\label{GllHres}
\end{equation}
\noindent
\begin{equation}
\bar{D}_l(x) = D_l(x) - O(1/L) \ ,
\label{DlHres}
\end{equation}
\noindent
\begin{equation}
\bar{G}_{x,x^{\prime}} = G_{x,x^{\prime}} + \frac{1}{2}\sum_{n=1}^2
(2\delta_{x_n,0}-1)(2\delta_{x^{\prime}_n,0}-1) \ .
\label{GxxHres}
\end{equation}
\noindent
One sees that the only new term which could potentially survive 
the TL is the second term in the last expression. This term
is to be substituted into (\ref{Q33}) and it leads
to the computation of sums of the form
\begin{equation}
P = \frac{1}{8}\sum_{b_1b_2}\sum_{x,x^{\prime}}
\sum_{i<j}^4\sum_{i^{\prime}<j^{\prime}}^4 
(4\delta_{x_1,0}\delta_{x^{\prime}_1,0}-4\delta_{x_1,0}+1) I_1 \ .
\label{Hcontr}
\end{equation}
\noindent
The first term vanishes like $O(1/L)$ because of 2 extra deltas.
The second and constant terms equal zero because of (\ref{canc}).
Moreover, in general it is clear from (\ref{GllHres})-(\ref{GxxHres}) 
that the holonomies may only contribute to the TL through
the constant term in $\bar{G}_{x,x^{\prime}}$.
This is, however equivalent to non-cancelation of infrared
divergences in some higher orders. In all other cases
the holonomies do not survive the TL.

\section{Summary and Discussion}

In this paper we used the link representation to develop
the low-temperature expansion of some $2D$ models.
Our main concern here was the investigation of a fixed distance
correlation function in models with non-abelian $SU(N)$ global
symmetry. We have calculated the first and second order coefficients 
of large-$\beta$ expansion of such correlation in terms of 
Green functions and have shown that they coincide with the
conventional expansion, at least for the shortest distance.
We have demonstrated how the path independence of 
the correlation function manifests itself in our expansion and 
have proven such independence for the first two coefficients
via direct calculations. We have also shown which properties
of the expansion guarantee its infrared finitness, at least
in lowest orders. Moreover, since in our expansion the source
of such divergences is exactly localized (they can only emerge
from the expectation values of auxiliary fields which appear
in even general power) we think it should be possible to
generalize the proof given in Section 4.1 for higher orders.
It could lead to a lattice analog of David's theorem \cite{IR}
which states the infrared finitness of the weak coupling
expansion of continuum models.

Further, we have re-analyzed the low-temperature expansion
of $1D$ models in the link representation. In this representation
one should impose a global constraint on link matrices
(\ref{jacob1D}). This global constraint vanishes in the TL
if this limit is taken before low-temperature expansion.
However, if expansion is done in a finite volume,
the expansion of the holonomy operator, which imposes
the global constraint does survive the TL. It leads
to the non-uniformity of the low-temperature expansion
in one dimension. We also have proven that it is not the case
in $2D$: at least in the lowest orders the holonomies do not
survive the TL and there is a good reason to believe
that they do not in higher orders as well.
 
We thus find that the low-temperature expansion performed
in the link representation coincides with the conventional PT,
at least in the lowest orders. In fact, it seems that these
two expansions have to coincide up to arbitrary order.
If the conventional PT produces the correct asymptotic expansion in 
a finite volume, any other expansion is bound to reproduce the same
coefficients when the volume is fixed. Moreover, this also refers 
to the TL if these coefficients are infrared finite.
Originally (following arguments given in \cite{su2link}) 
we thought that the expansion around $V_l=I$
has somehow better theoretical status. In particular, 
if the expansion around $U_x=I$ fails in the TL as the expansion
around the non-existing in the TL saddle point, 
there seems to be a little doubt that the point $V_l=I$ is
the only saddle point in the TL similarly to the abelian case.
It led us to a conclusion that the expansion in the link representation
could be free of ambiguities found in \cite{superinst}
(see also discussion in \cite{David}-\cite{QA}) and which 
put under doubts that the conventional PT gives an asymptotic
uniform in the volume for non-abelian models. 
Unfortunately, this our conclusion was wrong as the next example shows.
We consider this example as an superinstanton example \cite{superinst} 
adjusted for the periodic lattice in the link formulation.
Since the calculations are very lengthy
we only sketch the arguments leaving details for a future \cite{inprep}.

Let us consider the system on the periodic lattice
given by the partition function (\ref{lPF}) with the only difference:
we leave one plaquette, say $p_0$,  unrestricted, i.e. on the dual
lattice we have ($p_0\to x_0$)
\begin{equation}
J(V) = \prod_{x\ne x_0} 
\left[ \sum_r d_r \chi_r \left( \prod_{l\in p}V_l \right) \right] \ .
\label{jacobunr}
\end{equation}
\noindent
It has to be clear that this modification should leave no memory
in the TL. For example, it is the case in the high-temperature expansion
(truly, we checked only the lowest order).
We are going to argue that while the coefficients of the low-temperature
expansion of the $XY$ model are indeed unchanged in the TL,
it is not so for non-abelian models and the trouble starts
from $1/\beta^2$, precisely like in the superinstanton example.
The expansion is done precisely like with
the standard partition function, the only modification concerns
the generating functional. It can be shown that one has to use
the following Green functions
\begin{equation}
G_{ll^{\prime}} \to M_{ll^{\prime}} = G_{ll^{\prime}} +
\frac{1}{G_0}D_l(x_0)D_{l^{\prime}}(x_0) \ ,
\label{Gllun}
\end{equation}
\noindent
\begin{equation}
D_l(x) \to M_l(x) = D_l(x) - 
\frac{1}{G_0}G_{x,x_0}D_l(x_0) \ ,
\label{Dlun}
\end{equation}
\noindent
\begin{equation}
G_{x,x^{\prime}} \to M_{x,x^{\prime}} = G_{x,x^{\prime}} -
\frac{1}{G_0}G_{x,x_0}G_{x^{\prime},x_0} \ .
\label{Gxxun}
\end{equation}
\noindent
As the simplest examples, we give the expression for the average
plaquette in the $XY$ model up to the first order
\begin{equation}
\Gamma_{XY} (p=x) = 1 - \frac{1}{\beta}\frac{\delta_{x,x_0}}{G_0}  
+ O(\beta^{-2}) \ 
\label{XYpun}
\end{equation}
\noindent
and the expression for an average link of the unrestricted plaquette
up to the second order
\begin{equation}
\Gamma_{XY} (x_0,x_0+n) = 
1-\frac{1}{4\beta}(1+\frac{1}{4G_0}) - 
\frac{1}{32\beta^2}(1-\frac{1}{16G_0^2}) + O(\beta^{-3}) \ .
\label{XYcorresun}
\end{equation}
\noindent
One sees that even links of unrestricted plaquette converge to
the true TL, though the convergence is slow, like $O(1/\ln L)$
(similarly to the PT with superinstanton BC). 

The first order coefficient of the correlation function of 
the $SU(N)$ model also converges logarithmically to the TL value
(\ref{SUNC1}). However, it is not the case for the second order 
coefficient. This can be most easily seen, e.g. from the expression for
the function $M_l(x)$ in Eq.(\ref{Dlun}).
In the case when all sums converge to the TL uniformly, 
one gets in the TL the following expression
\begin{equation}
M_l(x) = D_l(x) - D_l(x_0) \ ,
\label{Dluntl}
\end{equation}
\noindent
which shows that there appear new contributions which do not 
vanish in the TL. We have not finished calculations of all
these contributions for links of restricted plaquette
but it seems for us that even in this case one should expect
a modification of the standard TL.

\end{document}